\documentclass[sigconf]{acmart}
\usepackage{amsmath,nccmath,tabularx} 
\usepackage{algorithm}
\usepackage{algorithmic}
\usepackage{color}
\usepackage{graphicx}
\graphicspath{{images/}}
\usepackage{wrapfig,lipsum}
\usepackage{tabularx}
\usepackage{graphicx}
\usepackage{lscape}
\usepackage{subcaption}
\usepackage{latexsym}
\usepackage{pifont}
\usepackage{multirow}

\usepackage{enumitem}
\setlist{leftmargin=3mm}
\usepackage[T1]{fontenc}
\usepackage[utf8]{inputenc}
\usepackage[font=small,labelfont=bf]{caption}
\usepackage{babel}

\usepackage{booktabs}

\usepackage{xcolor}
\usepackage{xspace}
\usepackage{bm}
\usepackage{amsmath}
\usepackage{placeins}

\AtBeginDocument{%
  \providecommand\BibTeX{{%
    \normalfont B\kern-0.5em{\scshape i\kern-0.25em b}\kern-0.8em\TeX}}}

\copyrightyear{2023} 
\acmYear{2023} 
\setcopyright{acmlicensed}\acmConference[SIGIR '23]{Proceedings of the 46th International ACM SIGIR Conference on Research and Development in Information Retrieval}{July 23--27, 2023}{Taipei, Taiwan}
\acmBooktitle{Proceedings of the 46th International ACM SIGIR Conference on Research and Development in Information Retrieval (SIGIR '23), July 23--27, 2023, Taipei, Taiwan}
\acmPrice{15.00}
\acmDOI{10.1145/3539618.3591952}
\acmISBN{978-1-4503-9408-6/23/07}

\begin{document}
\title[Augmenting Passage Representations with Query Generation]{Augmenting Passage Representations with Query Generation for Enhanced Cross-Lingual Dense Retrieval}



\author{Shengyao Zhuang}
\affiliation{%
	\institution{The University of Queensland}
	\city{Brisbane}
	\state{QLD}
	\country{Australia}}
\email{s.zhuang@uq.edu.au}

\author{Linjun Shou}
\affiliation{%
	\institution{Microsoft STCA}
	\city{Beijing}
	\country{China}}
\email{lisho@microsoft.com}

\author{Guido Zuccon}
\affiliation{%
	\institution{The University of Queensland}
	\city{Brisbane}
	\state{QLD}
	\country{Australia}}
\email{g.zuccon@uq.edu.au}


\begin{abstract}

Effective cross-lingual dense retrieval methods that rely on multilingual pre-trained language models (PLMs) need to be trained to encompass both the relevance matching task and the cross-language alignment task. However, cross-lingual data for training is often scarcely available. 
In this paper, rather than using more cross-lingual data for training, we propose to use cross-lingual query generation to augment passage representations with queries in languages other than the original passage language. These augmented representations are used at inference time so that the representation can encode more information across the different target languages. 
Training of a cross-lingual query generator does not require additional training data to that used for the dense retriever. The query generator training is also effective because the pre-training task for the generator (T5 text-to-text training) is very similar to the fine-tuning task (generation of a query). The use of the generator does not increase query latency at inference and can be combined with any cross-lingual dense retrieval method.
Results from experiments on a benchmark cross-lingual information retrieval dataset show that our approach can improve the effectiveness of existing cross-lingual dense retrieval methods. Implementation of our methods, along with all generated query files are made publicly available at  \url{https://github.com/ielab/xQG4xDR}.


\end{abstract}


\maketitle

\section{Introduction}
Pre-trained language model-based (PLM) dense retrievers (DRs) have achieved remarkable success in the task of English-only passage retrieval~\cite{karpukhin-etal-2020-dense,xiong2020approximate,zhan2020repbert,Zhan2021OptimizingDR,ren2021pair,ren2021rocketqav2,qu2021rocketqa,lin2021batch,lin2020distilling,hofstatter2020improving,hofstatter2021efficiently}. These models use a dual-encoder architecture that encodes both queries and passages with a PLM encoder into dense embeddings. They then perform approximate nearest neighbor (ANN) searching in the embedding space. Compared to traditional bag-of-words approaches, DRs benefit from semantic soft matching, which helps overcome the problem of word mismatch in passage retrieval~\cite{tonellotto2022lecture,zhao2022dense}.

To leverage the semantic modelling power of DRs, recent research has extended English-only DRs to support cross-lingual settings~\cite{asai2021one,sorokin2022ask,li2022learning,ren2022empowering,xorqa,longpre2021mkqa}, i.e, where queries and passages are in different languages. This is achieved using multi-lingual PLMs, such as multilingual BERT~\cite{DBLP:conf/naacl/DevlinCLT19}, in place of the English-only PLMs. This approach is particularly important in this setting where traditional bag-of-words methods are ineffective due to the limited number of matching terms across languages. In contrast, cross-lingual DRs (xDRs) are able to encode queries and passages in different languages into a shared embedding space, enabling efficient ANN search across languages. 
However, such multi-lingual PLM-based xDRs usually are less effective on the cross-lingual passage retrieval task than DRs in the English-only setting~\cite{xorqa}. 
The hypothesis to explain this result is that, in the English-only setting, a DR only needs to learn relevance matching between queries and passages. In contrast, a xDR not only needs to learn the relevance matching task, but also needs to learn how to align the embeddings of texts with similar semantic meaning but in different language~\cite{nair2022transfer,yang2022c3}. It is this language gap that makes cross lingual retrieval a relatively harder task for xDRs.

\begin{figure*}
	\centering
	\includegraphics[width=1\textwidth]{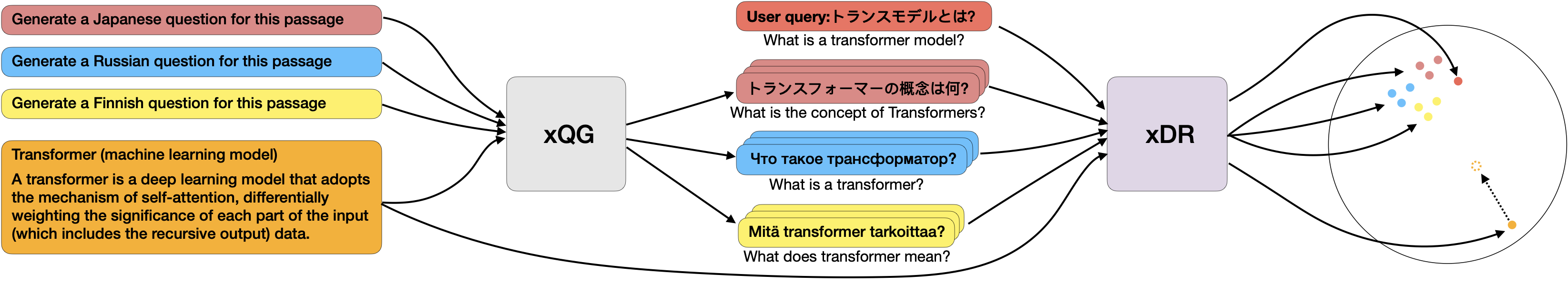}
	\vspace{-25pt}
	\caption{Augmenting passage representations with cross-lingual generated query embeddings. The query examples shown in this figure were generated using our trained xQG model. For each query, we report the corresponding translation obtained using Google's translation service. \vspace{-12pt}}
	\label{illustration}
\end{figure*}

Based on this hypothesis, this paper proposes the use of cross-lingual query generation (xQG) to bridge the language gap for xDRs. Our approach is illustrated in Figure~\ref{illustration}. Specifically, we fine-tune a multilingual T5 (mT5) model using language-specific prompts to generate several queries per passage for each target language. In the indexing phase, we use a given xDR model to encode passages and their associated generated queries in different languages into embeddings. Finally, we augment the original passage embeddings with the embeddings of the generated queries before adding them to the ANN index. By doing so, we move the passage embeddings to a space that is closer to the target user query language. Our approach does not add extra query latency and can be applied to any existing xDRs. 

\vspace{-10pt}
\section{Related works}

\textbf{\textit{Cross-lingual dense retrievers.}}
The development of cross-lingual pre-trained language models has significantly contributed to the progress of cross-lingual dense retrieval (xDR)~\cite{conneau2020unsupervised,wu2022unsupervised,chi2021infoxlm,feng2022language}. Notable examples of recent xDR methods include CORA~\cite{asai2021one}, which employs a generator to facilitate retrieval training data mining, Sentri~\cite{sorokin2022ask}, which proposes a single encoder and self-training, and DR.DECR~\cite{li2022learning}, which utilizes parallel queries and sentences for cross-lingual knowledge distillation. Among these, our work is most closely related to QuiCK~\cite{ren2022empowering}, which also utilizes a cross-lingual query generator for xDR. However, unlike QuiCK, which uses the xQG as a teacher model in knowledge distillation for xDR in the training phase, our method directly augments passage embeddings with xQG queries without any xDR training involved.

\textbf{\textit{Query generation for information retrieval.}} Query generation is a well-established technique that has been widely used to improve retrieval performance in various retrieval models~\cite{nogueira2019doc2query,gospodinov2023doc2query}. In addition, it has been shown to be effective for domain adaptation in dense passage retrieval tasks~\cite{thakur2021beir,ma2021zero,wang2022gpl}, as well as for enhancing the effectiveness of other PLM-based rankers~\cite{zhuang2022bridging,wangneural}. In our approach, we also rely on a query generation model to generate high-quality queries for downstream passage embedding augmentation tasks.

\textbf{\textit{Augmenting embeddings for dense retrievers.}} 
Our method relies on effectively augmenting representations encoded by DR encoders -- a direction recently explored also by other works. For instance, \citet{li2023pseudo,li2022pyserini} use the top retrieved passages as pseudo-relevant feedback to augment query embeddings using the Rocchio aggregation function. Similarly, \citet{zhuang2022implicit} extend this idea by using embeddings of clicked passages to augment query embeddings. Other PRF methods have also been extensively researched to enhance both English-only dense retrieval~\cite{yu2021improving,li2022improving,wang2021pseudo,wang2023colbert,zhu2022lol} and cross-lingual dense retrieval~\cite{chandradevan2022learning}. On the other hand, the HyDE method uses large pre-trained language models to generate hypothetical passages for a given query and directly uses the embedding of the hypothetical passage to perform the search~\cite{gao2022precise}. While all these works focus on augmenting query embeddings for DRs at query time, our work focuses on augmenting passage embeddings \textit{at indexing time}, thereby avoiding the extra overhead in terms of query latency.

\vspace{-8pt}
\section{Method}
\vspace{-2pt}
Our approach consists of two components: (1) a cross-lingual query generation model that, for each passage, generates high-quality queries in different languages, and (2) an embedding aggregation function that augments the original passage embedding with the embeddings of the generated queries.

To obtain a xQG model, we fine-tune a mT5 model with labeled relevant $(q_t, p)$ pairs, where $t$ is the target query language. We use a seq2seq objective akin to docTquery-T5's objective~\cite{nogueira2019doc2query}; in our case the input is the passage text with language-specific prompts:
\begin{equation}
	\textit{prompt }(t,p) = \textit{Generate a } [t] \textit{ question for this passage: } [p],
\end{equation}
where $[t]$ and $[p]$ are prompt placeholders for the target language and passage text, respectively. Once we have trained a xQG model, we can generate a set of queries for each target language and passage by replacing the placeholders accordingly:
\begin{equation}\small
	Q_p =\bigcup_{t\in T} Q_{p}^t;  \textit{ }Q_{p}^t=\textit{xQG }(\textit{prompt }(t,p)) \times n,
\end{equation}
where $T$ is the target language set and $Q_p$ is the set of all generated queries for passage $p$ which includes $n$ generated queries for each target language. In our experiments, we use a top-$k$ sampling scheme~\cite{fan2018hierarchical} to generate queries, and set $k=10$. We find that these simple language-specific prompts can effectively lead the T5 model to generate the desired output in that language.

For the embedding aggregation function, we use a Rocchio-like aggregation function that is similar to previous works that aggregated dense representations~\cite{zhuang2022implicit,li2023pseudo}. We use this aggregation function to obtain the embeddings of all passages in the corpus:
\begin{equation}
	emb(p, Q_p, \theta, \alpha) = (1-\alpha)\cdot\theta(p) + \alpha \sum_{\hat{q_t}\in Q_p}\theta(\hat{q_t}),
\end{equation}
where $\hat{q_t}$ is a generated query for target language $t$, $\theta$ is the xDR encoder model and $\theta(.)$ is the text embedding given by the xDR encoder. The hyper-parameter $\alpha$ is the augmentation ratio, $\alpha \in [0,1]$. This ratio is used to control the weights assigned to the original passage embedding and the generated query embeddings.

\begin{table*} [t]
	\centering
	\caption{xDR models trained with different backbone PLMs. Zero-shot means trained with only the English subset of the NQ training data (thus, zero-shot for the cross-lingual task). Statistical differences against the base model (without xQG) are labelled with $^\star$ ($p<0.05$). \vspace{-12pt}}
	\begin{subtable}[c]{1\linewidth}
		\centering
		\caption{R@2kt}   
		\vspace{-5pt}
		\scalebox{0.94}{
			\begin{tabular}{l|lllllll|l}
				\hline
				Model          & Ar    & Bn    & Fi    & Ja    & Ko    & Ru    & Te    & Average \\ \hline
				XLM-R       & 28.5 & 26.3 & 29.3 &  22.4 &  34.0 &  17.3 &  34.9 &   27.5 \\
				XLM-R + xQG & \textbf{30.1}  & \textbf{29.6} &  \textbf{31.2} & \textbf{23.7} & \textbf{36.1} &  \textbf{19.4} & \textbf{38.2} &  \textbf{ 29.8}$^\star$ \\ \hline
				mBERT            & 41.1 & 49.0 & 52.2 & \textbf{37.3}$^\star$ & 48.1& 33.3 & 47.9& 44.1  \\ 
				mBERT + xQG    & \textbf{42.4}& \textbf{54.9}$^\star$ & \textbf{54.1}& 33.6 & \textbf{52.3}$^\star$  & \textbf{33.8} & \textbf{52.5}$^\star$ & \textbf{46.2}$^\star$  \\ \hline
				mBERT (zero-shot)       &32.6  &  25.0 & 38.2 &  29.5 & 38.9 &  27.0 & 39.9  &  33.0  \\
				mBERT (zero-shot)  + XQG &  \textbf{33.9} &  \textbf{28.9}$^\star$ &  \textbf{43.0}$^\star$&  \textbf{32.8}$^\star$ &  \textbf{41.4} &  \textbf{29.5} &  \textbf{42.0} &   \textbf{36.0}$^\star$  \\ \hline
			\end{tabular}
		}    
	\end{subtable}\\
	\begin{subtable}[c]{1\linewidth}
		\centering
		\caption{R@5kt} 
		\vspace{-5pt}
		\scalebox{0.94}{
			\begin{tabular}{l|lllllll|l}
				\hline
				Model          & Ar    & Bn    & Fi    & Ja    & Ko    & Ru    & Te    & Average \\ \hline
				XLM-R       &  38.5 &  33.6 &  37.9 & \textbf{32.8} & 42.8 & 28.3 &  47.5 &    37.3 \\
				XLM-R + xQG &  \textbf{38.8} &  \textbf{41.1}$^\star$ &  \textbf{39.8} & \textbf{32.8} & \textbf{43.2} & \textbf{30.8}  &  \textbf{49.6} &  \textbf{39.4}$^\star$  \\ \hline
				mBERT             & 49.2& 57.6& \textbf{58.6}& 42.7 & 57.5 & \textbf{41.4} & 55.9 & 51.8  \\
				mBERT + xQG    & \textbf{51.5} & \textbf{60.9}$^\star$ & 58.3 & \textbf{43.6}& \textbf{58.6} & \textbf{41.4} & \textbf{60.1}$^\star$ & \textbf{53.5}$^\star$   \\ \hline
				mBERT (zero-shot)       & 38.5 & 36.5 & 47.5 &  38.2 &  48.1 & 35.0 & 48.7 &  41.8  \\
				mBERT (zero-shot)  + xQG & \textbf{43.7}$^\star$ & \textbf{40.8}$^\star$ &  \textbf{50.0}$^\star$ & \textbf{40.2}  &  \textbf{49.8} & \textbf{39.7}$^\star$ & \textbf{51.7} &  \textbf{45.1}$^\star$  \\ \hline
			\end{tabular}
		}
	\end{subtable}
	\label{main}
	\vspace{-14pt}
\end{table*}

\section{Experimental Settings}
We design our experiments to answer the research questions:
\begin{itemize}
	\setlength{\itemindent}{16pt}
	\item[\bf RQ1:] How does the number of generated queries affect xDR effectiveness?
	\item[\bf RQ2:] How does the augmentation ratio $\alpha$ impact xDR effectiveness?
	\item[\bf RQ3:] How do the queries generated for each language impact  xDR effectiveness?
\end{itemize}
\textbf{\textit{Datasets and Evaluation.}} We train and evaluate our approach on XOR-TyDi~\cite{xorqa}, a cross-lingual open retrieval question answering benchmark dataset. The dataset contains approximately 15k annotated relevant passage-query pairs in the training set and 2k passage-answer pairs in the dev set. Queries in both the train and dev sets are in seven typologically diverse languages (Ar, Bn, Fi, Ja, Ko, Ru, and Te), while passages are in English. There are about 18M passages in the corpus. We use the training set to train both the xQG model and the xDRs. We evaluate the effectiveness of our approach and baselines using recall at $m$ kilo-tokens (R@$m$kt), which is the dataset's official evaluation metric. This metric computes the fraction of queries for which the minimal answer is contained in the top $m$ tokens of the retrieved passages. We consider $m = 2k, 5k$ (R@2kt and R@5kt), as per common practice for this dataset. Statistical significant differences are reported with respect to a two-tailed t-test with Bonferroni correction. 

\textbf{\textit{Baselines.}} Following common practice on XOR-TyDi~\cite{xorqa}, we adapt DPR~\cite{karpukhin-etal-2020-dense} to the cross-lingual setting by initializing it with different multilingual pre-trained language models (PLMs). Specifically, in our experiments we use the multilingual variants of BERT~\cite{DBLP:conf/naacl/DevlinCLT19} (mBERT) and XLM-RoBERTa~\cite{conneau2020unsupervised} (XLM-R) for this purpose. In addition to the standard supervised baselines, we also explore how our xQG embedding augmentation approach can improve a zero-shot xDR model, where the xDR is initialized with mBERT but it is trained with English only passage-query pairs and is directly applied to the XOR-TyDi cross-lingual retrieval task.

\textbf{\textit{Implementation details.}} We initialize our xQG model with the multilingual T5-base checkpoint provided by Google\footnote{\url{https://huggingface.co/google/mt5-base}} and available in the Huggingface library~\cite{wolf-etal-2020-transformers}. We fine-tune this checkpoint with the passage-query pairs in the XOR-TyDi training set. We train the xQG model for 1600 steps using a batch size of 128 and a learning rate of 1e-4. After training, we generate 5 queries per passage for each target language, resulting in about 7 * 5 * 18M = 630M generated queries in total. We use four A100 GPUs to generate all queries; this process took about 70 hours to complete. We release our generated queries on the Huggingface hub~\footnote{\url{https://huggingface.co/datasets/ielab/xor-tydi-xqg-augmented}} to allow the community to reuse this resource~\cite{scells2022reduce}. For training the xDRs, we use the Tevatron dense retriever training toolkit~\cite{Gao2022TevatronAE}, which uses with BM25 hard negative passages. We train the xDRs with a batch size of 128, initializing them with mBERT base\footnote{\url{https://huggingface.co/bert-base-multilingual-cased}} or XLM-R base\footnote{\url{https://huggingface.co/xlm-roberta-base}} checkpoints and training them on the XOR-TyDi training set for 40 epochs. For each training sample, we set the number of hard negative passages in the contrastive loss to 7 and applied in-batch negatives training. We use a learning rate of 1e-5 for mBERT-based xDRs and of 2e-5 for XLM-R-based xDRs.
For the zero-shot xDR, we use the same training configurations as for the mBERT-based xDR trained on XOR-TyDi but using the Natural Questions (NQ) training data~\cite{47761} which contains English-only query-passage training samples.

\section{Results}
\subsection{Main results}
Table~\ref{main} presents the effectiveness of xDR models initialized with the XLM-R and mBERT backbone PLMs and trained on the XOR-TyDi dataset. Zero-shot denotes the models trained only on the English subset of the NQ dataset. In these experiments, we use all the queries generated by our xQG and set the augmentation ratio to $\alpha=0.01$ for augmenting the passage embeddings of the xDRs. 

\begin{figure}[t]
	\centering
	\includegraphics[width=0.85\columnwidth]{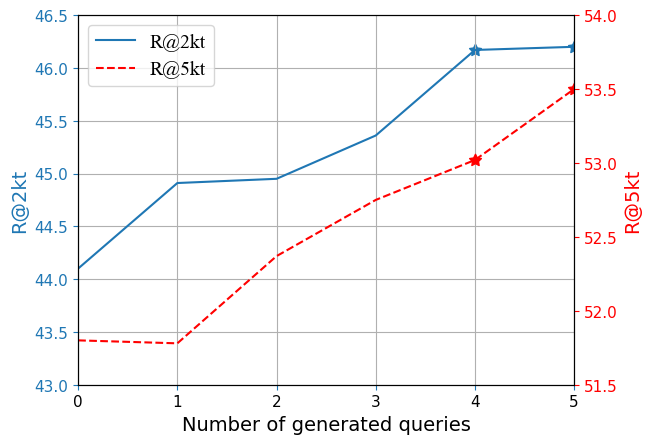}
	\vspace{-10pt}
	\caption{Impact of the amount of generated queries for target language. Scores are averaged across all languages. Statistical significant improvements over no augmentation (number of queries $n=0$) are labelled with stars ($p<0.05$). \vspace{-16pt}}
	\label{fig:num_query}
\end{figure}

\begin{figure}[t]
	\centering
	\includegraphics[width=0.83\columnwidth]{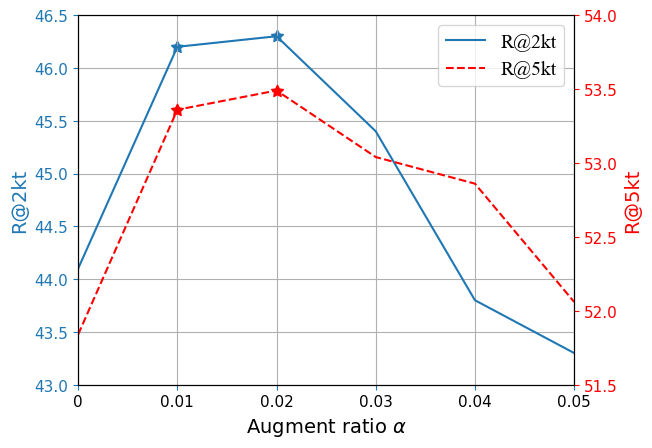}
	\vspace{-10pt}
	\caption{Impact of the augmentation ratio $\alpha$. Scores are averaged across all languages. Statistical significant improvements over no augmentation ($\alpha=0$) are labelled with stars ($p<0.05$). \vspace{-18pt}}
	\label{fig:ratio}
\end{figure}

For the R@2kt metric, the xDR initialized with mBERT outperforms the xDR initialized with XLM-R, achieving an average R@2kt  score of 44.1, while XLM-R achieves an average score of 27.5. 
Our xQG passage embedding augmentation approach improves the XLM-R xDR, achieving an average score of 29.8, which is a statistically significant improvement compared to its baseline effectiveness ($p < 0.05$). Similarly, mBERT's effectiveness improves with xQG, achieving an average score of 46.2, which is also a statistically significant improvement compared to its corresponding baseline ($p < 0.05$). The zero-shot mBERT model achieves an average R@2kt of 33.0; this also improves when combined with xQG, achieving an average score of 36.0. This improvement is statistically significant ($p < 0.05$).
Similar trends are found for R@5kt. 
Overall, we find that our xQG can significantly improve all investigated xDR models. 
In terms of per language effectiveness, xQG improves almost all models across all languages with the exceptions of mBERT's R@2kt for Japanese (Ja) and mBERT's R@5kt for Finnish (Fi).

In summary, mBERT performs better than XLM-R for both R@2kt  and R@5kt. The use of xQG embedding augmentation statistically significantly improves the effectiveness of both backbones.

\subsection{RQ1: Impact of number of generated queries}

Figure~\ref{fig:num_query} reports the impact of using different amounts of generated queries to augment passage embeddings when using mBERT xDR. The results suggest that using more generated queries is beneficial for both R@2tk and R@5tk. The improvements become statistically significant when 4 or more generated queries are used for each target language. While the curves do not plateau, indicating that using even more generated queries could further improve the effectiveness, our experiments were limited to up to 5 generated queries per target language due to computational constraints.

\subsection{RQ2: Impact of augmentation ratio}

We report the impact of the augmentation ratio $\alpha$ on the effectiveness of xDR in Figure~\ref{fig:ratio}. Higher values of $\alpha$ correspond to assigning more weight to the generated query embeddings during embedding aggregation, and $\alpha=0$ corresponds to no augmentation. As shown in the figure, even a small value of $\alpha$ (0.01) leads to a significant improvement in both R@2kt and R@5kt. The best effectiveness is achieved when $\alpha=0.02$. However, using higher values of $\alpha$ does not result in further improvements -- rather, it can even hurt the effectiveness when $\alpha>0.05$. Based on these results, we conclude that the augmentation ratio in our embedding aggregation function has a significant impact on the effectiveness of xDR, and using a small value of $\alpha$ can be beneficial for improving effectiveness.

\subsection{RQ3: Impact of each languages}

\begin{figure}[t]
	\centering
	\includegraphics[width=1\columnwidth]{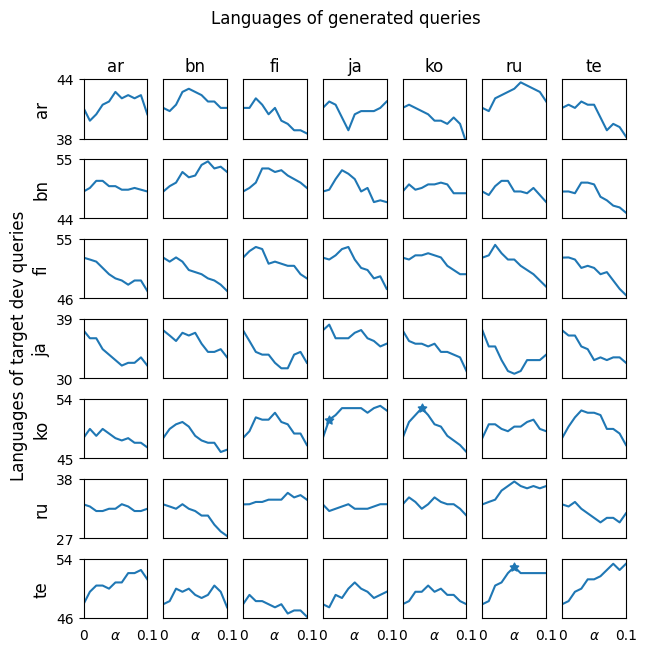}
	\vspace{-16pt}
	\caption{R@2kt for target languages (rows) augmented by source generated queries (columns). We plot $\alpha$ from 0 to 0.1 with step size of 0.01 (x axis). Statistically significant better results with respect to no augmentation ($\alpha=0$) are labelled with stars ($p<0.05$). \vspace{-14pt}}
	\label{fig:per_lang}
\end{figure}

In the previous experiments we used all the queries generated for a passage to augment the original passage embedding, irrespective of language of the generated query. Next, we investigate the impact on effectiveness of using generated queries from each of the languages separately. 
We analyze this in Figure~\ref{fig:per_lang}, where we plot R@2kt for each target language (rows) against different values of $\alpha$ (x-axis). 

The plots in the diagonal show that xDR effectiveness improves when passage embeddings are augmented with generated queries for the same language.
  Notably, we observe that the best value of $\alpha$ varies across different languages, and the improvements in effectiveness are not always statistically significant. We also observe that, for some languages, queries generated for other languages can improve the effectiveness of target queries in a different language. For instance, using queries generated for Japanese (Ja) can improve the effectiveness of target queries in Korean (ko), while using Russian (Ru) generated queries can help target queries in Telugu (Te).
These results suggest that the embeddings of the queries generated for any single language potentially can also provide useful information for target queries in other languages.

\section{Conclusion and future work}

In this paper we propose a passage embedding augmentation approach to enhance the effectiveness of cross-lingual DRs. Our method can be applied to any cross-lingual DR and it requires no further DR training nor changes to the retrieval pipeline. We empirically showed the method is effective for the cross-lingual DPR method across different backbones. However, a limitation of our empirical investigation is that we did not evaluate the method across other dense retriever architectures. We leave the extension of this investigation to future work. We also note that our approach relies on a xQG model that can generate high quality queries. However, a recent work has shown that T5-based query generation is prone to hallucination and that along with highly effective queries, the generator also produces poor queries that negatively impact retrieval effectiveness~\cite{gospodinov2023doc2query}. They then propose the use of a cross-encoder ranker to filter out some ineffective generated queries; this practice can further improve  effectiveness. We leave the adaptation of this approach in our xQG setting to future work.



\bibliographystyle{ACM-Reference-Format}
\balance
\bibliography{sigir2023}

\appendix

\end{document}